\documentclass{article}

\usepackage{arxiv}
\usepackage{moreverb}
\usepackage{amsmath}

\usepackage{natbib}
\usepackage{hyperref}       
\usepackage{url}  
\usepackage{amsfonts}       
\usepackage{nicefrac}       
\usepackage{microtype}      
\usepackage{lipsum}		
\usepackage{graphicx}
\usepackage{doi}

\newcommand\BibTeX{{\rmfamily B\kern-.05em \textsc{i\kern-.025em b}\kern-.08em
T\kern-.1667em\lower.7ex\hbox{E}\kern-.125emX}}

\providecommand{\bo}{\mathbf}
\providecommand{\bs}{\boldsymbol}

\DeclareMathOperator*{\argmin}{argmin}

\title{Reducing bias and alleviating the influence of excess of zeros with multioutcome adaptive LAD-lasso}

\renewcommand{\shorttitle}{}
\author{ \href{https://orcid.org/0000-0002-6270-2556}{\includegraphics[scale=0.06]{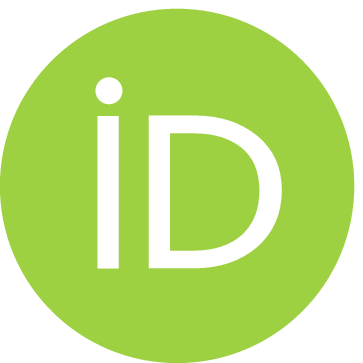}\hspace{1mm}Jyrki Möttönen}\thanks{Corresponding author} \\
	Department of Mathematics and Statistics\\
	University of Helsinki\\
	Helsinki, Finland\\
	\texttt{jyrki.mottonen@helsinki.fi} \\
	\And
	\href{https://orcid.org/0000-0001-5856-2899}{\includegraphics[scale=0.06]{orcid.eps}\hspace{1mm}Tero Lähderanta} \\
	Research Unit of Mathematical Sciences\\
	University of Oulu\\
	Oulu, Finland \\
	\And
	\href{https://orcid.org/0000-0002-0595-6226}{\includegraphics[scale=0.06]{orcid.eps}\hspace{1mm}Janne Salonen} \\
	Research Department \\
	Finnish Centre for Pensions and \\
	Keva \\
	Helsinki, Finland \\ 
	\And
	\href{https://orcid.org/0000-0003-2808-2768}{\includegraphics[scale=0.06]{orcid.eps}\hspace{1mm}Mikko J. Sillanpää} \\
	Research Unit of Mathematical Sciences\\
	University of Oulu\\
	Oulu, Finland
	
}

\begin{document}
\maketitle

\begin{abstract}
	Zero-inflated explanatory variables are common in fields such as ecology and finance. In this paper we address the problem of having excess of zero values in some explanatory variables which are subject to multioutcome lasso-regularized variable selection. Briefly, the problem results from the failure  of the lasso-type of shrinkage methods to recognize any difference between zero value occurring either in the regression coefficient or in the corresponding value of the explanatory variable. This kind of confounding will obviously increase number of false positives - all non-zero regression coefficients do not necessarily represent real outcome effects.

We present here the adaptive LAD-lasso for multiple outcomes which extends the earlier work of multivariate LAD-lasso with adaptive penalization. In addition of well known property of having less biased regression coefficients, we show here how the adaptivity improves also method's ability to recover from influences of excess of zero values measured in continuous covariates.
\end{abstract}

\keywords{Multivariate analysis; $p>n$ regression; Penalized regression; Robust Procedures; Variable Selection}



\maketitle

\section{Introduction}
\label{sec:intro}

In high-dimensional regression problems the number of parameters is often larger than the number of individuals in the sample. Ordinary least squares and other conventional estimation methods don’t work in such situations. Simultaneous estimation and variable selection using lasso \citep{tibshirani1996,li2012} is a popular shrinkage-estimation approach to obtain sparse estimates for regression coefficients in high-dimensional regression problems. However, the generally known drawback of shrinkage-inducing methods, including lasso, is that they improve estimation accuracy but also introduce downward bias to the estimates. To alleviate this problem, a re-estimation procedure has been proposed where effects of selected predictors are re-estimated using no penalty \citep{efron2004,meinshausen2007}. Adaptive lasso has been also presented \citep{zou2006}, so that the selected predictors are subject of reduced penalty and unselected positions obtain heavier penalty. This is performed either by using two-stage or iterative estimation strategy. See also related work of thresholded lasso \citep{zhou2010,vandegeer2011}.

When the outcome distributions are skewed or they contain some outlying observations, it is possible to apply robust LAD regression instead of ordinary regression. In lasso context, LAD-lasso has been proposed for univariate \citep{wang2007} and multioutcome cases \citep{mottonen2015,li2015}. The multioutcome LAD-lasso is related to the group lasso \citep{yuan2006} because the object function of multioutcome LAD-lasso also contains group lasso penalty. As common shrinkage-inducing methods, also LAD-lasso methods are suffering from downward bias of the estimates. To alleviate this, adaptive LAD-lasso has been proposed for univariate LAD-lasso \citep{arslan2012}.
In this paper we study the multioutcome LAD-lasso. Note that adaptive group lasso has been presented in \citet{wang2008}.

We have pinpointed a potential problem in LAD-lasso or lasso method when there are excess of real zero values in some covariates. We describe this problem here and provide also some potential solutions. Shortly, the problem occurs as the lasso-type of shrinkage methods do not recognize any difference between zero value occurring either at the regression coefficient or in the corresponding value of the covariate. Now there is a danger of misinterpretation as  non-zero regression coefficients do not necessarily all represent real effects.
As a potential solution, we show here how use of the adaptive LAD-lasso will minimize this kind of confounding to happen in practice.  

\section{Multioutcome LAD-lasso}
\label{sec:LAD-lasso}

Consider multiple regression model with multiple outcomes
$$
\bo Y = \bo X\bo B + \bo E,
$$
where $\bo Y = (\bo y_1,\ldots,\bo y_n)'$ is an $n\times p$ matrix of $n$ observed values of $p$ outcome variables, $\bo X = (\bo x_1,\ldots,\bo x_n)'$ is an $n\times q$ matrix of $n$ observed values of $q$ explanatory variables, $\bo B$ is a $q\times p$ matrix of regression coefficients, and $\bo E = (\bs\varepsilon_1,\ldots,\bs\varepsilon_n)'$ is an $n\times p$ matrix of residuals. We further assume that $\bs\varepsilon_1,\ldots,\bs\varepsilon_n$ is a random sample of size $n$ from a $p$-variate distribution centered at the origin.

The multivariate lasso estimation method is based on  
the penalized objective function
\begin{eqnarray}
\label{lasso2}
\frac1n\sum_{i=1}^n\|\bo y_i-{\bo B}'\bo x_i\|^2 + \lambda\sum_{j=2}^q\|\bs\beta_j\|
\end{eqnarray}
(See e.g. \citet{turlach2005}, \citet{yuan2006} and \citet{yuan2007}). The minimizer of the objective function (\ref{lasso2}) gives now {\it multivariate lasso estimate} for the regression coefficient matrix $\bo B$. Note that equation \ref{lasso2} is also an objective function of group lasso where each of  $q-1$ explanatory variables (the intercept terms are omitted) have $p$ regression coefficients (one for each response variable) which form an group.

The multivariate lasso method gives sparse solutions but it is obviously not very robust. You will get a more robust version by minimizing the penalized objective function (where the squared norms are replaced with norms) 
\begin{equation}
\label{lad-lasso3}
\frac1n\sum_{i=1}^n\|\bo y_i-{\bo B}'\bo x_i\|
+ \lambda\sum_{j=2}^q\|\bs\beta_j\|
\end{equation}
with respect to the coefficient matrix $\bo B$ \citep{mottonen2015,li2015}. Denote the minimizing value of (\ref{lad-lasso3}) by $\hat{\bo B}_{LL}$. It is easily seen that  if we define
$$
\begin{pmatrix}
\bo y_i^* \\
\bo x_i^*
\end{pmatrix}
=
\left\{
\begin{array}{ll}
\begin{pmatrix}\bo y_i\\ \bo x_i\end{pmatrix}, & \text{if\ }\ i = 1,\ldots,n,\\
\\
\begin{pmatrix}\bs 0\\ n\lambda\bo e_{i-n+1}\end{pmatrix}, & \text{if\ }\ i = n+1,\ldots,n+q-1,
\end{array} \right.
$$
the objective function (\ref{lad-lasso3}) reduces to the LAD estimation objective function \citep{oja2010}
\begin{equation}
\label{lad-lasso4}
\frac1{n+q-1}\sum_{i=1}^{n+q-1}\|\bo y_i^*-{\bo B}'\bo x_i^*\|,
\end{equation}
which shows that we can use any multioutcome LAD regression estimation routine to find the multioutcome LAD-lasso estimate $\hat{\bo B}_{LL}$. You can, for example, use the function {\it mv.l1lm} of the R-package {\it MNM} \citep{nordhausen2016,nordhausen2011}. 

\vfill\eject

\section{Multioutcome adaptive LAD-lasso}
\label{sec:adaptive-LAD-lasso}

The multioutcome LAD-lasso estimate for a fixed $\lambda$ can be defined as
$$
\hat{\bo B}^*(\lambda)
=
\argmin_{\beta}\left[
\frac1n\sum_{i=1}^n\|\bo y_i-{\bo B}'\bo x_i\|
+ \lambda\sum_{j=2}^q\|\bs\beta_j\|
\right].
$$

The question then arises how to choose the tuning parameter $\lambda$. If we are mainly concerned about recovering the right model, then you can use, for example, Akaike's information criterion (AIC) or Bayesian information criterion (BIC) \footnote{For a BIC-like criterion in multioutcome LAD-lasso context, see \citet{mottonen2015}.}. On the other hand, if we are mainly concerned about prediction accuracy, then cross-validation technique is often a good choice.

Let
$$
\lambda_0^* = \argmin_{\lambda}\ \text{CV}(\hat{\bo B}^*(\lambda))
$$
be the value of the tuning parameter $\lambda$ which
minimizes the cross-validation criterion (or alternatively AIC or BIC criterion) function for the multioutcome LAD-lasso. The multioutcome LAD-lasso estimate can then be defined as
$$
\hat{\bo B}^* = \hat{\bo B}^*(\lambda_0^*).
$$
It has been shown that lasso-estimation tends to underestimate the regression coefficients and the same is true for the multioutcome LAD-lasso estimate $\hat{\bo B}^*$.\ \ 
\citet{zou2006} proposed an adaptive lasso method  which gives an estimate whose bias is smaller than that of the standard lasso. The adaptive method can also be used in the LAD-lasso \citep{arslan2012}. We extend this method here for multioutcome LAD-lasso case.  The multivariate adaptive LAD-lasso estimate for tuning parameters $\lambda_j$, $j=2,\ldots,q$ can be defined as
\begin{eqnarray*}
\hat{\bo B}(\lambda)
& = &
\argmin_{\beta}\left[
\frac1n\sum_{i=1}^n\|\bo y_i-{\bo B}'\bo x_i\|
+ \sum_{j=2}^q\lambda_j\|\bs\beta_j\|
\right],
\end{eqnarray*}
where
$$
\lambda_j = \frac{\lambda}{\|\bs\beta_j^*\|+1/n},\ \ j=2,\ldots,q
$$
and $\bs\beta_j^*$ is the $j$th row of the multioutcome LAD-lasso estimate $\hat{\bo B}^*={\hat{\bo B}}^*(\lambda_0^*)$. If we denote $w_j^*=(\|\bs\beta_j^*\|+1/n)^{-1}$, then the adaptive multioutcome LAD-lasso estimate can be written in the form
\begin{eqnarray*}
\hat{\bo B}(\lambda)
& = &
\argmin_{\beta}\left[
\frac1n\sum_{i=1}^n\|\bo y_i-{\bo B}'\bo x_i\|
+ \lambda\sum_{j=2}^qw_j^*\|\bs\beta_j\|
\right]
\\
& = &
\argmin_{\beta}\left[
\frac1n\sum_{i=1}^n\|\bo y_i-{\bo B}'\bo x_i\|
+ \frac1n\sum_{j=2}^q\|\bo 0-{\bo B}'n\lambda w_j^*\bo e_j\|
\right]
\end{eqnarray*}
which further implies that it can be written in the multioutcome LAD regression form
\begin{equation}
\label{ad-lad-lasso3}
\hat{\bo B}(\lambda)
=
\argmin_{\beta}\left[
\frac1{n+q-1}\sum_{i=1}^{n+q-1}\|\bo y_i^*-{\bo B}'\bo x_i^*\|
\right],
\end{equation}
where
$$
\begin{pmatrix}
\bo y_i^* \\
\bo x_i^*
\end{pmatrix}
=
\left\{
\begin{array}{ll}
\begin{pmatrix}\bo y_i\\ \bo x_i\end{pmatrix}, & \text{if\ }\ i = 1,\ldots,n,\\
\\
\begin{pmatrix}\bo 0\\ n\lambda w_j^*\bo e_{i-n+1}\end{pmatrix}, & \text{if\ }\ i = n+1,\ldots,n+q-1,
\end{array} \right.
$$
Let
\begin{equation}
\lambda_0 = \argmin_{\lambda}\ \text{CV}(\hat{\bo B}(\lambda))
\label{CV_criterion}
\end{equation}
be the value of the tuning parameter which minimizes the cross-validation criterion function for the estimate $\hat{\bo B}(\lambda)$. {\it Adaptive multioutcome LAD-lasso estimate} is then 
$$
\hat{\bo B}={\hat{\bo B}}(\lambda_0).
$$ 
Since the performance of the adaptive LAD-lasso might be sensitive to the initial weights $\gamma_j$, the following iterative estimation procedure can give more stable results:

\begin{itemize}
\item[(S1)]
\ \ \ Find the initial multioutcome LAD-lasso estimate 
${\hat{\bo B}}^* = \hat{\bo B}(\lambda_0^*)$, where $\lambda_0^*$ minimizes the cross-validation
criterion function \ref{CV_criterion}.
\item[(S2)]
\ \ \ Calculate
$
w_j^* = (\|\bs\beta_j^*\|+1/n)^{-1}, \ \ j=2,\ldots,q.
$
\item[(S3)]
\ \ \ Find ${\hat{\bo B}} = \hat{\bo B}(\lambda_0)$, where $\lambda_0$ minimizes the
cross-validation criterion function \ref{CV_criterion}.
\item[(S4)]
\ \ \ Calculate
$
w_j = (\|\bs\beta_j\|+1/n)^{-1}, \ \ j=2,\ldots,q.
$
\item[(S5)]
\ \ \ Iterate through steps (S3)-(S4) until convergence measured as 
$$
\frac{\|\hat{\bo B}^{(s)}-\hat{\bo B}^{(s-1)}\|}
     {\|\hat{\bo B}^{(s-1)}\|}
$$
is obtained.    
\end{itemize}

\section{Simulation studies}

In this section we present two simulated data sets to provide insights to the multioutcome adaptive LAD-lasso in certain scenarios. R-code for the simulations can be found on github \citep{Lahderanta2021}.

\subsection{Bias reduction}
\label{sec:simulation}

For this study, we simulated new phenotypes with trivariate traits using the public genotype data set from the 12th QTL-MAS workshop in Uppsala, Sweden, in 2008. The original genotype data set contains 5,865 individuals and 6,000 markers. We took a random sample of size 300 individuals and chose every 30th marker with a resulting total number of 200 markers. We then generated trivariate traits by using the multivariate multiple regression model
$$
\bo Y = \bo X\bo B + \bo E,
$$
where $\bo Y$ is a $300\times 3$ matrix of trivariate traits, $\bo X$ is a $300\times200$ matrix with $ij$th element  
$$
x_{ij}
=
\left\{
\begin{array}{rl}
-1, & \text{if indiv. $i$ is homozygote (11) at marker $j$},\\
 0, & \text{if indiv. $i$ is heterozygote at marker $j$},\\
 1, & \text{if indiv. $i$ is homozygote (22) at marker $j$},
\end{array} \right.
$$

\begin{figure}[!ht]
\centerline{\includegraphics[width=11cm,height=9.78cm]{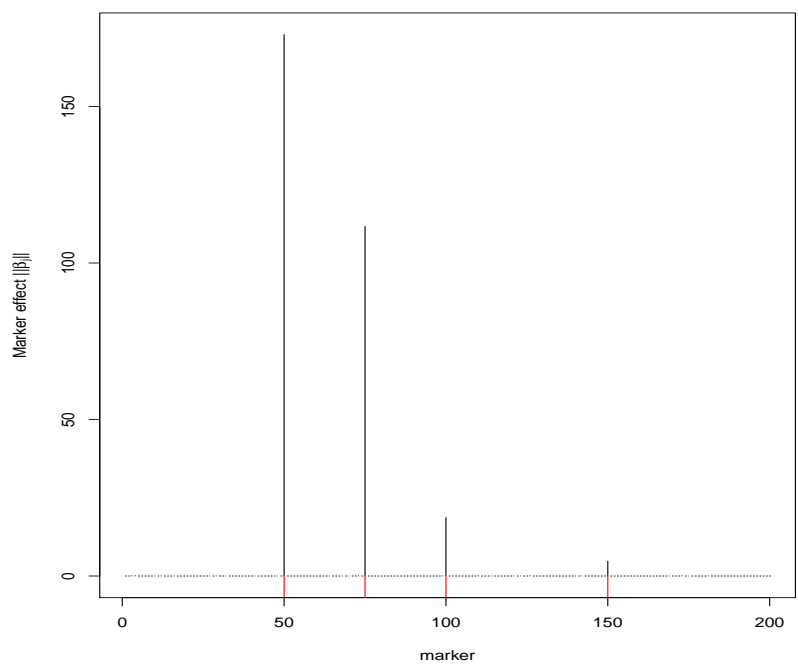}}
\caption{\label{fig1}Marker effects of the multioutcome LAD-lasso estimates. The red ticks show the locations of the four QTLs (markers 50, 75, 100 and 150).}
\end{figure}

$\bo B$ is a $200\times 3$ matrix with four QTLs indicated as non-zero rows
\begin{eqnarray*}
\bs\beta_{50}' & = & (100,100,100),\ \ \bs\beta_{75}'=(0,50,100) \\
\bs\beta_{100}' & = & (5,10,15)\ \ \mbox{and}\  \ \bs\beta_{150}'=(3,3,3),
\end{eqnarray*}
and $\bs\varepsilon$ is a $300\times 3$ matrix with i.i.d. rows distributed as
$$
N_3\left(
\begin{pmatrix}
0  \\ 
0  \\ 
0 
\end{pmatrix},    
\begin{pmatrix}
1.0 & 0.5 & 0.3 \\ 
0.5 & 1.0 & 0.2 \\ 
0.3 & 0.2 & 1.0
\end{pmatrix} 
\right).
$$
We estimated the tuning parameter $\lambda$ by using 5-fold cross-validation. The Figure \ref{fig1} shows the marker effects $\|\bs\beta_j\|$, $j=1,\ldots,200$.  We see that the LAD-lasso method correctly finds all four QTLs.

\begin{figure}[!ht]
\centerline{\includegraphics[width=11cm,height=9.78cm]{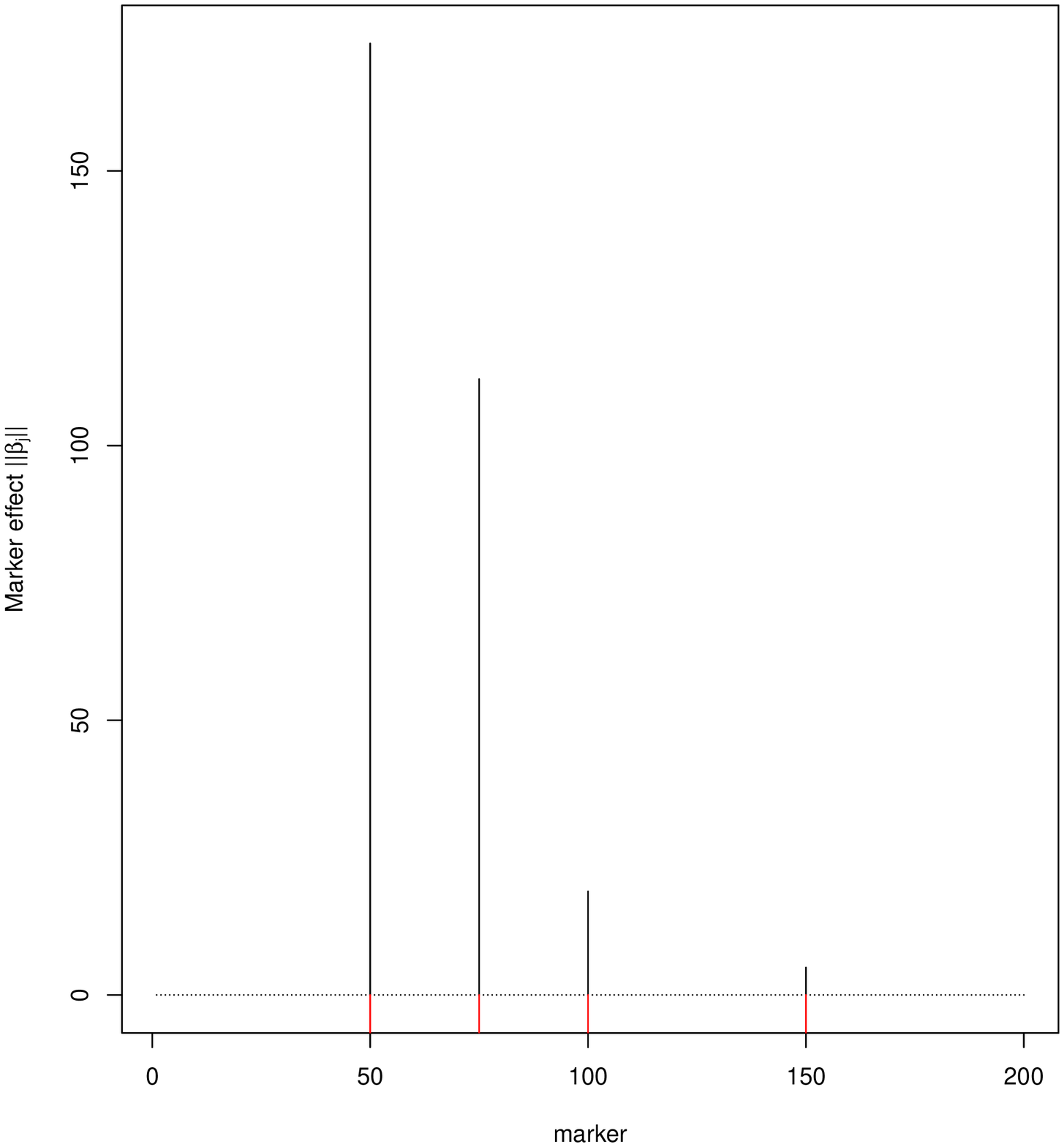}}
\caption{\label{fig2}Marker effects of adaptive multioutcome LAD-lasso estimates. The red ticks show the locations of the four QTLs (markers 50, 75, 100 and 150).}
\end{figure}

We then studied the bias of the LAD-lasso estimates of the non-zero coefficient vectors ${\bs\beta}_{50}$, ${\bs\beta}_{75}$, ${\bs\beta}_{100}$ and ${\bs\beta}_{150}$:
\begin{eqnarray*}
\begin{pmatrix}
\hat{\bs\beta}_{50}'-{\bs\beta}_{50}'\\
\hat{\bs\beta}_{75}'-{\bs\beta}_{75}'\\
\hat{\bs\beta}_{100}'-{\bs\beta}_{100}'\\
\hat{\bs\beta}_{150}'-{\bs\beta}_{150}'
\end{pmatrix}
& = &
\begin{pmatrix}
-0.15 & -0.19 & -0.08 \\
-0.09 & -0.08 & -0.13 \\
~~0.00 & ~~0.04 & -0.12 \\
-0.29 & -0.22 & -0.34
\end{pmatrix}
\end{eqnarray*}
We see that the LAD-lasso estimates are noticeably biased.

Then we used the adaptive multioutcome LAD-lasso estimation method. The Figure \ref{fig2} shows the marker effects $\|\bs\beta_j\|$, $j=1,\ldots,200$. We see that the adaptive multioutcome LAD-lasso method correctly finds the QTLs.

\begin{figure}[!ht]
\centerline{\includegraphics[width=11cm,height=9.78cm]{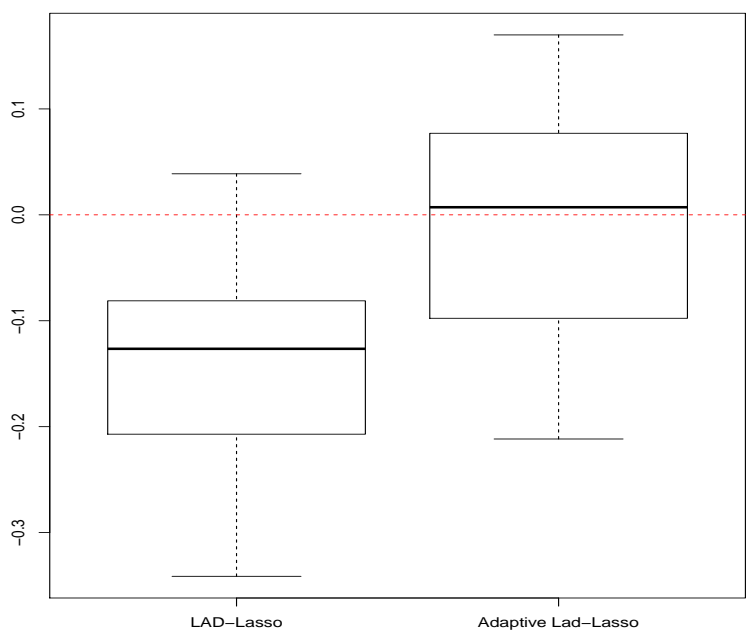}}
\caption{\label{fig3}Box-plots of the biases of the regression coefficient estimates}
\end{figure}

The biases of the non-zero coefficient vectors were in  this case
\begin{eqnarray*}
\begin{pmatrix}
\hat{\bs\beta}_{50}'-{\bs\beta}_{50}'\\
\hat{\bs\beta}_{75}'-{\bs\beta}_{75}'\\
\hat{\bs\beta}_{100}'-{\bs\beta}_{100}'\\
\hat{\bs\beta}_{150}'-{\bs\beta}_{150}'
\end{pmatrix}
& = &
\begin{pmatrix}
-0.06  & -0.10  & ~~0.01 \\
~~0.00 & ~~0.12 & ~~0.17 \\
~~0.03 & ~~0.15 & ~~0.02 \\
-0.17  & -0.09 & -0.21 
\end{pmatrix}
\end{eqnarray*}
We can see that the estimates are now less severely biased. The Figure \ref{fig3} shows that the biases of the adaptive regression estimates are scattered around zero but the non-adaptive regression estimates are scattered around -0.15.

\begin{figure}[!ht]
\centerline{\includegraphics[width=11cm,height=9.78cm]{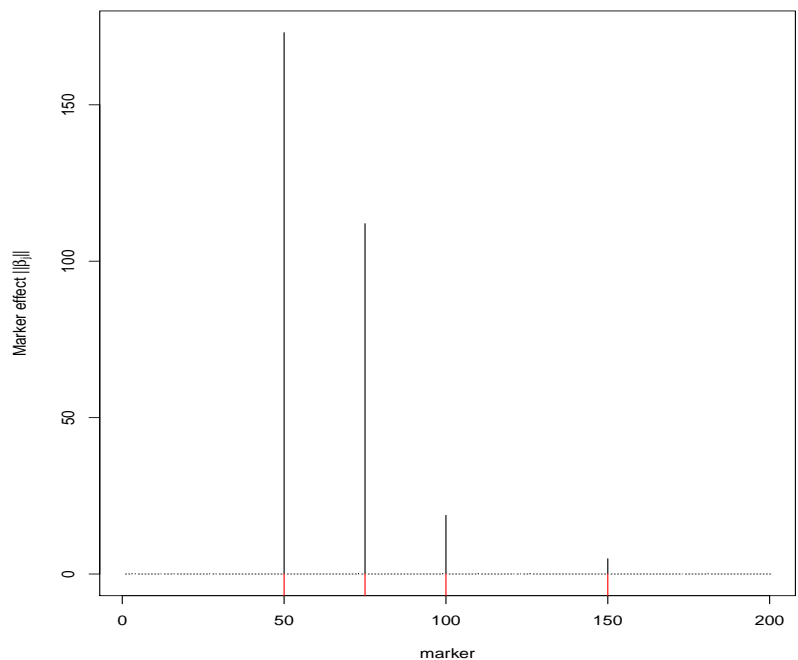}}
\caption{\label{fig4}Marker effects of adaptive multioutcome LAD-lasso estimates for the contaminated data. The red ticks show the locations of the four QTLs (markers 50, 75, 100 and 150).}
\end{figure}

We then constructed a simple robustness study. We multiplied $\bo y_{10}$ and $\bo y_{292}$ by 100 and calculated the adaptive multioutcome LAD-lasso estimates. The Figure \ref{fig4} shows that the adaptive  method finds the QTL's also in the contaminated data case.

The biases of the regression coefficients are now
\begin{eqnarray*}
\begin{pmatrix}
\hat{\bs\beta}_{50}'-{\bs\beta}_{50}'\\
\hat{\bs\beta}_{75}'-{\bs\beta}_{75}'\\
\hat{\bs\beta}_{100}'-{\bs\beta}_{100}'\\
\hat{\bs\beta}_{150}'-{\bs\beta}_{150}'
\end{pmatrix}
& = &
\begin{pmatrix}
-0.08 & -0.12  & -0.00\\
-0.02 &   ~~0.15  &  ~~0.16\\
~~0.01  &  ~~0.11  &  ~~0.00\\
-0.20 &  -0.11  & -0.23
\end{pmatrix}
\end{eqnarray*}

which indicates that the outliers had only a minor effect on the biases.

\subsection{Excess of zeros}

In this study, we simulate a different data set to demonstrate the excess of zeros scenario with multioutcome adaptive LAD-lasso and regular LAD-lasso. 

\begin{figure}[ht]
\centerline{\includegraphics[scale = 0.7]{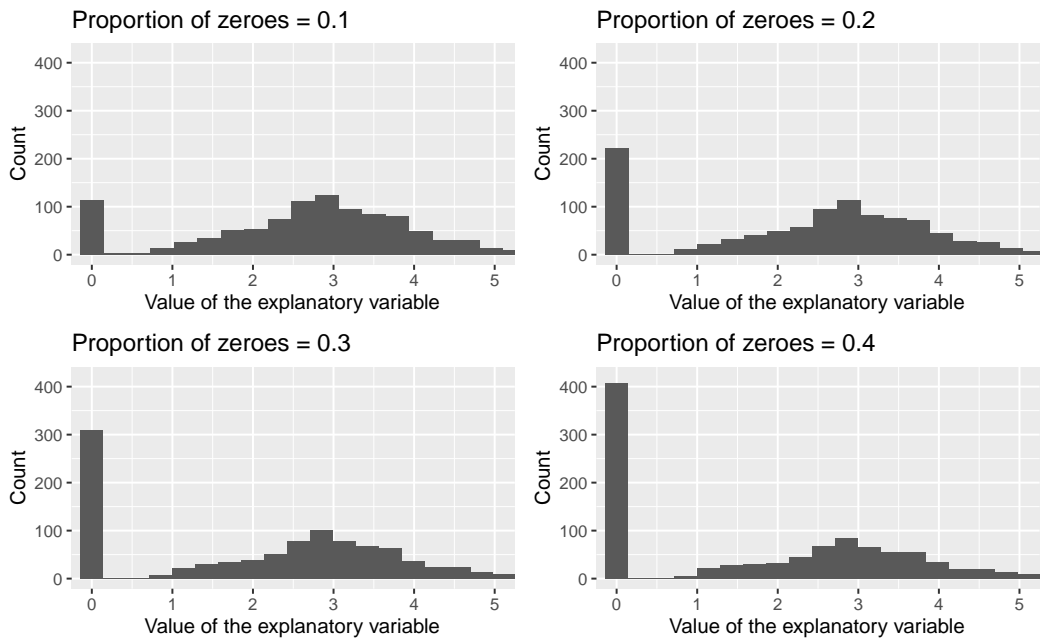}}
\caption{\label{figdatexm}Simulated data replicates in the excess of zeros study with different proportions of zeros.}
\end{figure}

Multiple data sets are shown to illustrate the performance of algorithms in wide variety of situations. We generate two types of data sets, one with high number of observations compared to covariates ($n = 100, q = 10$), and contrarily one with high number of covariates when compared to observations ($n = 25, q = 50$). Moreover, we alternate the proportion of zeros in the covariates from 0.1 to 0.4 and apply two different error $\bo E$ distributions: uniform and asymmetric Laplace. When all the combinations of these properties are consider, in total of 16 different types of scenarios are examined. In each scenario we simulate 100 data sets to further evaluate the impact of the methods. The data sets and the alternating proportion of zeros is shown in Figure \ref{figdatexm}.

The data sets are generated with multioutcome regression model  
$$
\bo Y = \bo X{\bo B} + \bo E,
$$

with matrix $\bo X$ of a size $n\times q$, response matrix $\bo Y$ of a size $n\times 2$. The  observed values $x_{ij}$ are simulated from a normal distribution such as
$$
x_{ij}
\sim
\left\{
\begin{array}{rl}
N(3,1), & \text{if $u_{ij} > p_{zeros}$},\\
0, & \text{else},
\end{array} \right.
$$

where $u_{ij} \sim Unif(0,1)$ and $p_{zeros}$ is the proportion of zeros.

$\bo B$ is a $q\times 2$ matrix where 

$$
b_{jk}
\sim
\left\{
\begin{array}{rl}
N(0,1), & \text{when $j = 1,2,3$},\\
0, & \text{else}.
\end{array} \right.
$$

$\bo E$ is a $n\times 2$ matrix where 

$$
e_{jk}
\sim
Unif(0,1)
$$ or

$$
e_{jk}
\sim
ALaplace(\mu = [3,6],\Sigma = 
\begin{pmatrix}
0.5  & 0 \\
0 & 0.5  
\end{pmatrix}),
$$
depending of the choice for error distribution. Above, "ALaplace" stands for Asymmetric Laplace distribution.

\begin{figure}[ht]
\centerline{\includegraphics[scale = 0.7]{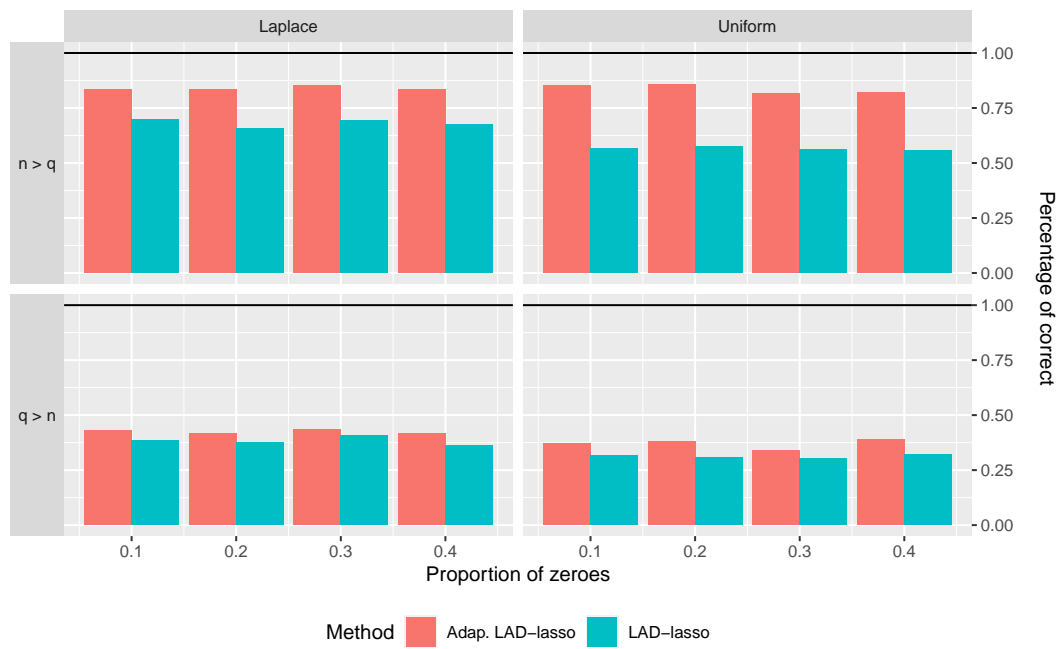}}
\caption{\label{figsim}Percentage of correctly found zero coefficients with adaptive LAD-lasso and regular LAD-lasso in multioutcome context.  }
\end{figure}

From the simulated data sets, we can observe that the adaptive LAD-lasso is superior to non-adaptive in every scenario, when we compare the correctly found zero coefficients (Figure \ref{figsim}). In scenario $q > n$ the difference between the methods is much smaller. 

\section{Concluding remarks}
\label{sec:concluding}
We have shown here that the multioutcome adaptive LAD-lasso is a versatile robust tool and is capable of reducing bias from effect estimates as well as alleviating the influence from the problem of excess of zero values in continuous covariates. The competitive performance of adaptive version compared to multioutcome LAD-lasso has been illustrated with several examples. In the future, it would be interesting to develop a robust BIC criterion for multivariate LAD-lasso context, in order to be able to account for linear dependencies between outcomes.

\section{Acknowledgments}
\label{sec:acknowledgments}

This research is supported by the Infotech Oulu spearhead project funding.

\vfill\eject


\bibliographystyle{unsrtnat}
\bibliography{references}

\end{document}